\documentclass[prl,twocolumn,amsmath,amssymb,
superscriptaddress,floatfix]{revtex4-1}

\usepackage{graphicx}
\usepackage{color}
\usepackage{float}
\usepackage[normalem]{ulem}
\usepackage{hyperref}
\usepackage{bm}

\newcommand{\bk}{\bm{k}}

\newcommand{\bp}{\bm{p}}

\newcommand{\be}[0]{\begin{equation}}
\newcommand{\ee}[0]{\end{equation}}
\newcommand{\ba}[0]{\begin{eqnarray}}
\newcommand{\ea}[0]{\end{eqnarray}}

\newcommand{\up}[0]{\uparrow}
\newcommand{\dn}[0]{\downarrow}
\newcommand{\bmat}[0]{\begin{bmatrix}}
\newcommand{\emat}[0]{\end{bmatrix}}

\begin{document}
\title{Kondo destruction and fixed-point annihilation in a Bose-Fermi Kondo model}

 \author{Haoyu Hu}
\email{hh25@rice.edu}
\affiliation{Department of Physics \& Astronomy, Rice Center for Quantum Materials, Rice University, Houston, Texas 77005, USA}
\author{Qimiao Si}
\email{qmsi@rice.edu}
\affiliation{Department of Physics \& Astronomy, Rice Center for Quantum Materials, Rice University, Houston, Texas 77005, USA}

\begin{abstract}
Quantum
criticality that goes beyond the Landau framework of order-parameter fluctuations
is playing a central role
in elucidating the behavior of strange metals. A prominent 
case appears
in Kondo lattice systems, 
which have been extensively analysed in terms of an effective Bose-Fermi Kondo model. Here, a spin is
simultaneously coupled to conduction electron bands and 
gapless vector bosons that represent magnetic fluctuations. 
The Bose-Fermi Kondo model
 features interacting fixed points of Kondo destruction
  with such properties as dynamical Planckian ($\hbar \omega / k_{\rm B} T$) scaling and 
loss of quasiparticles.
Here we carry out a renormalization-group analysis 
of the 
model with spin isotropy
and identify pair-wise annihilations of the fixed points
as the spectrum of the bosonic bath evolves.
Our  analysis not only provides an 
essentially complete understanding of the previous numerical results of an SU(2)-symmetric model, 
but also reveals a surprising feature of sequential fixed-point annihilation.
Our results lay the foundation for the understanding of quantum criticality in spin-isotropic heavy-fermion metals 
as well as in doped Mott-Hubbard systems.

\end{abstract}

\maketitle

\textit{Introduction: }
Quantum criticality appears in a variety of strongly correlated metallic systems
\cite{QCNP2013,Coleman-Nature,Pas21.1,Kir20.2,Sachdev-book}, 
with the antiferromagnetic (AF)  case being prototypical.
It provides a setting for strange metals with a complete loss of quasiparticles and 
with a dynamical Planckian ($\hbar \omega /k_{\rm B} T$) scaling
\cite{Si-Nature,Colemanetal,senthil2004a}. 
Such properties are to be contrasted with the Landau description of metallic
quantum criticality.
In the Landau framework, phases of matter are 
identifieded
by order parameters, which specify the spontaneous breaking of global symmetries,
and quantum criticality is described by the 
spatial and temporal fluctuations of the order parameter:
A continuous AF to paramagnetic phase transition at  $T=0$ is characterized by
 the slow fluctuations 
of the staggered magnetization \cite{Hertz, Millis}; 
at such a 
spin-density-wave 
QCP,
quasiparticles remain and the underlying fixed point is Gaussian,
leading to a violation of dynamical Planckian scaling. 
In AF heavy-fermion metals, the beyond-Landau description
\cite{Si-Nature,Colemanetal,senthil2004a} features the added critical ingredient in the form of Kondo destruction.
Experiments in many 
heavy fermion strange metals
are in strong support for
the beyond-Landau nature
of the quantum critical points
 \cite{Pas21.1,Kir20.2,paschen2004,Gegenwart2007,Friedemann.10,shishido2005,park-nature06,Kne08.1,Custers-2012,Mar19.1,Schroder,Prochaska20,shot-noise2022}.

Given the broad current interest in the strange metal behavior \cite{Kei17.1,Pas21.1,Phillips22},
it is opportune to elucidate its physics
by expanding our understanding of the
 Kondo destruction
 quantum criticality.
The Bose-Fermi Kondo (BFK) model 
 has been extensively recognized as an important case study in this context,
given that it is  an effective description \cite{doi:10.7566/JPSJ.83.061005} of the Kondo lattice Hamiltonian
  in
  the extended dynamical mean-field theory (EDMFT) \cite{Si.96,SmithSi-edmft,Chitra}.
The model has been treated by renormalization group (RG) method within
an $\epsilon$-expansion
{ (see below for the definition of $\epsilon$)}
\cite{si1996kosterlitz,smith1999,sengupta2000,Si-Nature,si2003.prb,zhu2002,zarand2002}.
Various numerical means
ranging from Monte-Carlo to numerical renormalization group 
methods
\cite{GrempelSi,ZhuGrempelSi,Glossop07,Zhu07}
have been used to analyse the Ising-anisotropic model.
For the SU(2)-symmetric
BFK model, numerical studies have only recently become  possible:
Ref.\,\onlinecite{Cai_2019} developed a continuous-time Monte Carlo method for this case and
 identified an unexpectedly large number (eight in total) of
 fixed points 
for small (but positive) $\epsilon$, 
which are pair-wise annihilated as $\epsilon$ is increased.
Understanding these 
surprising  results
is important not only
for the beyond-Landau quantum criticality and strange metal physics 
but also for the decoherence physics of 
spin qubits.

In this work, we carry out analytical calculations to identify all the fixed points
of the SU(2)-symmetric BFK model
seen 
in the numerical simulations.
Our calculation utilizes an RG method that is
based on a large-$S$ expansion (with $S$ denoting the spin size)~\cite{cuomo2022spin}.
Our results not only capture the phenomena of pair-wise fixed-point annihilations as a function of increasing $\epsilon$,
as suggested by the numerical results, but also uncover a hitherto unsuspected 
result that such annihilations develop in sequence. 
The BFK model can be understood as an effective $0+1$ dimensional nonlinear $\sigma$ model, 
but with not only a topological term but also an extra Kondo coupling to gapless fermions. 
As we show later, the Kondo coupling 
affects the topological term.
The interplay among the Kondo interaction, topological term, and bosonic coupling 
leads to a rich phase diagram with interacting fixed points that annihilate in sequence.
Our results are important for the understanding of quantum criticality in 
correlated systems such 
as the physics of Kondo destruction in heavy fermion metals 
and doped Mott-Hubbard systems.

\textit{Model: }
We focus on the following
 Hamiltonian  for the BFK model:
\ba 
H &=& J_K \sum_{\mu,a} S^\mu s^\mu_{c,a}(\bm{r_0}) 
+ g \sum_\mu S^\mu \phi^\mu(\bm{r_0}) \nonumber \\
&+&   \sum_{\bk,\sigma,a}\epsilon_{\bk} c_{\bk,\sigma,a}^\dag c_{\bk,\sigma,a}+\frac{1}{2}
\sum_{\bp,\mu} E_{\bp}\phi_{-\bp}^\mu \phi_{\bp}^\mu \, .
\label{eq:ham}
\ea 
Here, $S^{\mu}(\mu\in\{x,y,z\})$ denotes a spin-$S$ impurity at site $\bm{r_0}=\bm{0}$. $c_{\bk,\sigma,a}^\dag$ 
creates a conduction electron with momentum $\bk$, spin $\sigma$, orbital $a$ (where $a\in \{1,2,..,K\}$) 
and dispersion $\epsilon_{\bk}$. 
We consider a generic filling of the conduction electrons with a Fermi surface in the non-interacting limit, 
which gives a constant conduction electron density of state near the Fermi energy: 
$\rho_c(\epsilon) = \sum_{\bk}\delta (\epsilon - \epsilon_{\bk}) \sim \rho_0$.
The spin operator of the conduction electron with orbital $a$ at site $\bm{r_0}$ is defined as $s_{c,a}^\mu(\bm{r_0}) = c_{\bm{r_0},a}^\dag \frac{\sigma^\mu }{2}c_{\bm{r_0},a}$,
where
 $\sigma^{x,y,z}$ 
are the
 Pauli matrices. It couples to the impurity at the same site via Kondo coupling $J_K$.
$\phi_{\bp}^\mu$ describes a free $O(3)$ bosonic field with momentum $\bp$ and kinetic energy $E_{\bp} = \bp^2$.
The spectral functions of $\phi$ fields are taken to be
$\rho_b(\omega) = \sum_{\bp}\delta(\omega-E_{\bp})\propto |\omega|^{1-\epsilon}$.
$g$ is the coupling strength between bosonic field at site $\bm{r}_0$ and impurity spin.
To perform a large $S$ expansion, we also rescale the fields and coupling constants as
$\phi^\mu \rightarrow \sqrt{S}\phi^\mu$, $g\rightarrow S^{-1/2}g$, $J_K\rightarrow 2S^{-1}J_K$. 
All the terms of the Hamiltonian are
then of order $S$. 
Throughout the work, we  focus on the case of perfect screening with $\kappa=K/S=2$.

\textit{RG analysis and beta functions:} 
We now carry out an RG analysis of the model defined in Eq.~\ref{eq:ham}. 
To do so, we first calculate the propagators of the fermions and bosons using the large $S$ expansion.

We note that these propagators can be expressed as functionals of the $n$-point correlation functions of $S^\mu$ [see Supplementary Material 
 (SM) A~\cite{SM}]. Thus, it's sufficient to consider the effective theory of $S^\mu$ and its correlation functions. The 
effective action of $S^\mu$ reduces to 
\ba 
&&
S_{eff}
 \nonumber \\
&=&
 -iSS_B -\frac{g^2}{2} \int_{\tau,\tau'} \sum_{\tau,\tau'}S^\mu(\tau)  
G_\phi(\tau-\tau') S^\mu(\tau')\nonumber \\
&&
-K\text{Tr}\log\bigg[ \delta_{\tau,\tau'}\bigg(\delta_{k,k'}(\partial_{\tau'}  + \epsilon_{\bk}) 
+ J_K\sum_\mu S^\mu(\tau)\frac{\sigma^\mu}{2}\bigg)\bigg]\nonumber 
\, ,
\\
\label{eq:action}
\ea 
where $S_B$ describes the Berry phase for the spin,
the kernel $G_\phi(\tau) \propto |\tau-\tau'|^{-(2-\epsilon)}$
is obtained by integrating out the bosonic fields and the last line represents the contribution from integrating out the conduction electrons. 
We then introduce the spinor representation of the impurity spin $S^\mu =S z_a^\dag \frac{\sigma_{ab}^\mu}{2}z_b$ with a constraint,
 $|z_1|^2+|z_2|^2=2$~\cite{auerbach2012interacting}. 
 We expand the action around the saddle point solution $z=z_0 = \text{const}$ and 
integrate out zero-modes to enforce the $SU(2)$ 
symmetry~\cite{zinn2007phase}. 
Without loss of generality, we consider $z = z_0 + \delta z$ with 
\ba 
z_0=\sqrt{2} \begin{bmatrix}1 & 0 \end{bmatrix}\, ,  ~~~~~~~~
\delta z = \begin{bmatrix} -\frac{|\chi|^2}{\sqrt{2}S} &  \frac{i\chi^\dag}{\sqrt{S}} \end{bmatrix} \, .
\label{eq:z_field}
\ea 
Here, the $\chi$ fields parametrize the $1/S^0$ fluctuations around the saddle-point solution
 $z_0$~\cite{auerbach2012interacting}. 
 The effective action at order $S^0$ is 
\ba 
&&S_{\chi} = \int  G^{-1}_\chi(i\Omega)|\chi(i\Omega)|^2 \frac{d\Omega}{2\pi}  \, ,
\label{eq:thy_chi}
\ea 
with the propagator
(see
SM  A~\cite{SM}):
\ba 
G_\chi^{-1}(i\Omega) &= &i\Omega \bigg[ 1+p(J|\Omega|^{-r})\bigg] +|\Omega|\bigg[ q(J|\Omega|^{-r}) + h|\Omega|^{-\epsilon}\bigg] \nonumber
\,  . \\
\label{eq:prop}
\ea 
Here, the functions $q(x)$ and $p(x)$ are defined as  
\ba 
&&q(x) =
2\kappa x^2 \,/ \left [\pi \, (1+x^2)^2  \right ] \, ,
 \nonumber \\
&&p(x) = 
(\kappa / \pi )
[x-x^3- (1+x^2)^2\arctan(x)]/
(1+x^2)^2
\nonumber 
\, ,
\ea 
where $p(x) \sim -8\kappa x^3/(3\pi)$ at small $x$,
and $p(x\rightarrow \infty) \rightarrow -\kappa/2 $.
In the derivation, we regularize the conduction electron density of state to be $\rho_c(\epsilon) = \rho_0|\epsilon|^{-r}$. 
The exponent $r$ is introduced to perform a minimal-subtraction 
RG study and will be set to zero at the final step of the calculation~\cite{zhu2002}. 
Other constants in Eq.~\ref{eq:prop} are $h \propto g^2$, $J = \pi J_K \rho_0$. 

With the effective theory of $\chi$, which describes the $S^0$ fluctuations, 
we're able to calculate the correlation functions of  the fermions and bosons (see SM A~\cite{SM}). 
We then require that the poles in these correlation functions are minimally removed~\cite{zhu2002,zinn2007phase}, 
which gives the following beta functions 
(see SM A~\cite{SM}):
\ba 
\beta_J &=&  \frac{J-J^3}{2S\pi (1+J^2)}F(J) + \frac{J^2}{S\pi(1+J^2)}H(J) \nonumber \\
&&+\bigg[ \frac{-(1-J^2)Jh+2iJ^2h}{2S(1+J^2)\pi c(J)(c(J)+h)}+\text{h.c.}
\bigg] +o(\frac{1}{S}) \nonumber  \, , \\
\beta_h& =& -\frac{1}{S}\tilde{\epsilon} h + \frac{h}{S\pi}F(J)
-\frac{1}{ S\pi }\frac{(c(J)+c(J)^*)h}{|c(J)|^2} 
\nonumber \\
&&-\frac{1}{S\pi }\frac{-2h^2-(c(J)+c(J)^*)h}{(c(J)+h)(c(J)^*+h)} +o(\frac{1}{S})  \, .
\label{eq:beta_eq}
\ea 
Here,
\ba 
F(j) &=& 
{2q(j)} / {\left [ (1+p(j))^2+q(j)^2 \right ]}
\nonumber \\ 
H(j) &= &
{-2(1+p(j))}/ { \left [ (1+p(j))^2 + q(j)^2 \right ] }
\nonumber \\
c(J) &= & i(1+p(J)) + q(J) ,
\ea 
and we rescale $\epsilon$ as $\epsilon = \tilde{\epsilon}/S$. 
All terms in the beta functions at the order of $1/S$ have been captured.
By treating $1/S$ as the only small parameter in the beta function, we're able to 
determine the phase diagram even for
large $J$ and $h$. 

\begin{figure*}
    \centering
    \includegraphics[width=\textwidth]{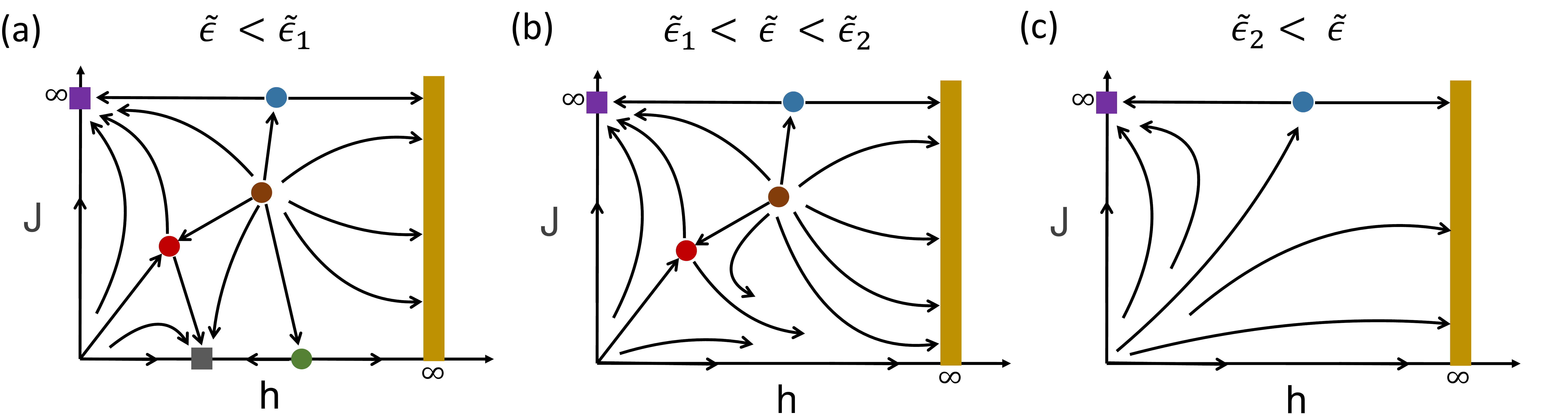}
    \caption{Schematic renormalization group flows for $\tilde{\epsilon}<\tilde{\epsilon}_1$ (a), $\tilde{\epsilon}_1<\tilde{\epsilon}<\tilde{\epsilon}_2$ (b) and $\tilde{\epsilon}_2<\tilde{\epsilon}$ (c). The squares label stabled fixed points, the circles mark unstable fixed points and the wide line denotes a line of stable fixed points at $h\rightarrow \infty$. See 
    SM C
     for the RG flow from an explicit numerical evaluations of the beta functions. }
    \label{fig:rg_flow}
\end{figure*}

\textit{RG flow: }
We are now in a position to analyze the RG flow. 
Note that there is always a trivial
unstable fixed point at $J=h=0$, which 
we will not discuss further. {In addition, at $J=0$, the model becomes a Bose-Kondo model and 
our results are consistent with previous works~\cite{cuomo2022spin,nahum2022fixed}.}
After numerical evaluations of Eq.~\ref{eq:beta_eq} at perfect screening $\kappa=2$, we identify three different types of phase diagrams corresponding to three regions of $\tilde{\epsilon}$: $\tilde{\epsilon}<\tilde{\epsilon}_1 $, $\tilde{\epsilon}_1<\tilde{\epsilon}<\tilde{\epsilon}_2$, $\tilde{\epsilon}_2<\tilde{\epsilon}$, with $\tilde{\epsilon}_1=1/\pi,\tilde{\epsilon}_2\simeq 0.36$. 
The schematic RG flows are shown in Fig.~\ref{fig:rg_flow}. There are in total seven types of non-trivial fixed points with their 
properties 
{-- including their labels --}
 listed in Tab.~\ref{tab:fixpoint}. 

We first analyze the RG flow at $\tilde{\epsilon} <\tilde{\epsilon}_1 $ as shown in Fig.~\ref{fig:rg_flow} (a). 
At $h=0$, the model is equivalent to a standard Kondo impurity model and the behavior of $\beta_J$ at small $J$ (see 
SM C~\cite{SM}) is consistent from what has been found via perturbation theory~\cite{affleck1995conformal}. 
Numerically, we find $J$ is always relevant and the system flows to a strong coupling fixed point at $J=\infty$. We call it Kondo or K fixed point;
the bosonic coupling $h$ is irrelevant at this fixed point. For $J=0$, 
there are 
 two stable fixed points at $h = (1-\sqrt{1-\tilde{\epsilon}^2})/ {\tilde{\epsilon}},\infty$ 
 and one critical point at $h=(1+\sqrt{1-\tilde{\epsilon}^2})/ {\tilde{\epsilon}}$. 
Due to their local moment nature, we label the two stable fixed points as L fixed point (at small $h$) and L$'$ fixed point (at infinity $h$). We call the critical point that controls the quantum phase transition between L and L$'$ as LC. After introducing the Kondo coupling, we find $J$ is irrelevant at both L and LC, but marginal at L$'$. The marginal behavior extends to finite $J$ which yields a line of fixed points at $h=\infty$ (see 
SM 
C~\cite{SM}). 

\begin{table}[H]
\centering
    \begin{tabular}{c|c|c|c | c}
    \hline 
     & Position & Stable? & Region & Color  \\
     \hline 
     L& $J=0,h=\frac{1-\sqrt{1-\tilde{\epsilon}^2}}{\tilde{\epsilon}}$&  Stable & $\tilde{\epsilon}<\tilde{\epsilon}_1$ & Grey \\
     \hline 
     L$'$ &$ h=\infty$& Stable &  & Yellow \\
     \hline 
     K & $J=\infty,h=0$ & Stable & & Purple \\ 
     \hline 
     LC & $J=0,h=\frac{1+\sqrt{1-\tilde{\epsilon}^2}}{\tilde{\epsilon}} $ & Unstable  &$\tilde{\epsilon}<\tilde{\epsilon}_1$ & Green\\
     \hline 
     KD & $J\ne 0,h\ne 0$ & Unstable &$\tilde{\epsilon}<\tilde{\epsilon}_2$ &Red\\
     \hline 
     KD$'$ & $J=\infty,h=\frac{2}{\tilde{\epsilon}\pi}$ & Unstable & &Blue\\
     \hline 
     C & $J\ne 0,h\ne 0$ &Unstable &$\tilde{\epsilon}<\tilde{\epsilon}_2$ & Brown\\
     \hline 
    \end{tabular}
\caption{List of fixed points with their positions in the phase diagram, stability, region of existence and color in Fig.~\ref{fig:rg_flow}}
\label{tab:fixpoint}
\end{table}

Away from $J=0$, $h=0$ and $h=\infty$ lines, there are three unstable fixed points. 
We label them as KD, KD$'$, and C. Both KD and KD$'$ describe the Kondo destruction quantum phase transition 
with one relevant direction and one irrelevant direction. A third fixed point C with all the directions being relevant separates 
KD and KD$'$ in the RG flow. 
KD is located at smaller $J,h$ and controls the Kondo destruction quantum phase transition between the local momentum phase L 
and Kondo-screened phase K. 
The quantum phase transition between L$'$ and K is described by the fixed point KD$'$ 
at $(h_c=\frac{2}{\tilde{\epsilon}\pi},J=\infty)$ (see
SM C~\cite{SM}). 

The topology of RG flow and the existence of fixed points change as we increase the value of $\tilde{\epsilon}$. 
Remarkably, we find  a two-step sequence
 of fixed-point annihilation.
 The first annihilation occurs between L and LC$'$ at $\tilde{\epsilon}=\tilde{\epsilon}_1=1/\pi$ and the second fixed-point annihilation occurs between C and KD at $\tilde{\epsilon}=\tilde{\epsilon}_2\simeq 0.36$. We show the RG flow after first annihilation in Fig.~\ref{fig:rg_flow} (b) and the flow after second annihilation in Fig.~\ref{fig:rg_flow} (c). 

\begin{figure*}
    \centering
    \includegraphics[width=0.9\textwidth]{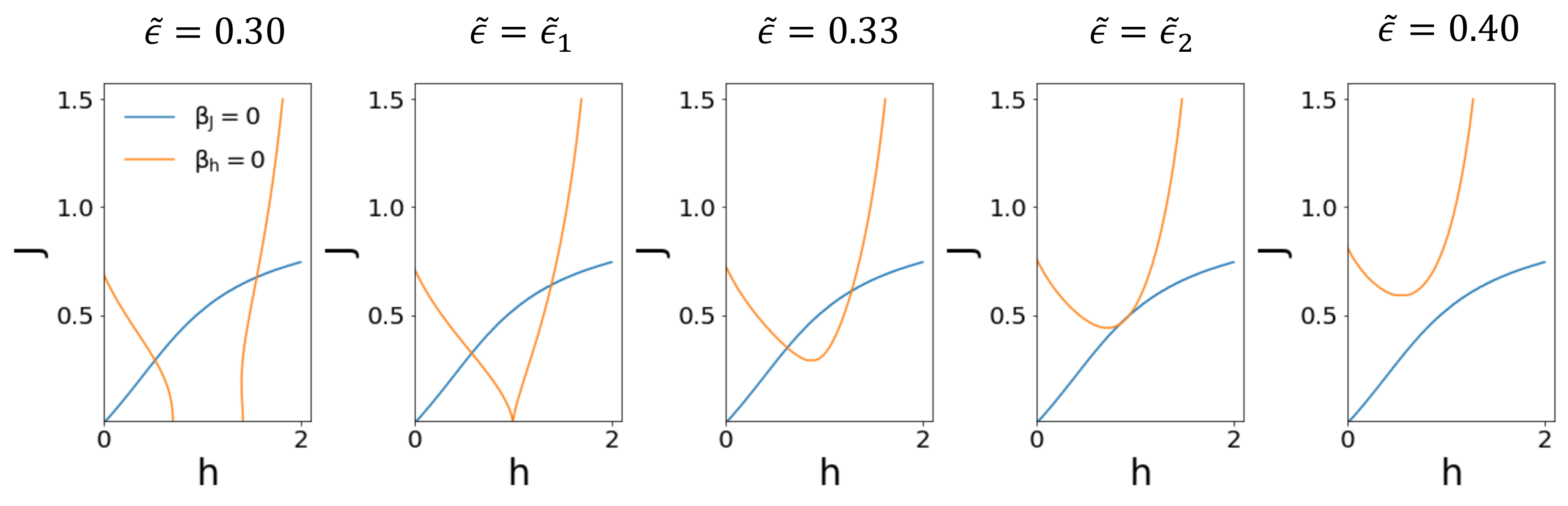}
    \caption{Zeros of $\beta_J$ and $\beta_h$ at $\tilde{\epsilon}<\tilde{\epsilon}_1$, $\tilde{\epsilon}=\tilde{\epsilon}_1$, $\tilde{\epsilon}_1<\tilde{\epsilon}<\tilde{\epsilon}_2$, $\tilde{\epsilon}=\tilde{\epsilon}_2$ and $\tilde{\epsilon}>\tilde{\epsilon}_2$. $\beta_h=0$ line have two crossing points with $J=0$ at $\tilde{\epsilon}<\tilde{\epsilon}_1$, which describe L and LC fixed points respectively. Two crossing points merge into one at $\tilde{\epsilon}=\tilde{\epsilon}_1$ and disappear at $\tilde{\epsilon}>\tilde{\epsilon}_1$. $\beta_h=0$ and $\beta_J=0$ lines have two crossing points at $\tilde{\epsilon}<\tilde{\epsilon}_2$, which represent KD and C fixed points respectively. They merge into one at $\tilde{\epsilon}=\tilde{\epsilon}_2$ and disappear at $\tilde{\epsilon}>\tilde{\epsilon}_2$.  }
    \label{fig:zero}
\end{figure*}

\textit{Fixed-point annihilation:} 
We next turn to a detailed analysis of the fixed-point annihilation. The annihilation between two fixed points 
{respectively} with 
$n$
{and $n+1$}
 relevant directions
 is generically allowed by the topology of the RG flow. 
In the BFK model, both $n=0$ and $n=1$ types of annihilation happen. To 
be specific,
 we consider the zeros of beta functions. 
The shape of $\beta_J=0$ line and $\beta_h=0$ line at various $\tilde{\epsilon}$ is shown in Fig.~\ref{fig:zero}. At $\tilde{\epsilon}<\tilde{\epsilon}_1$, $\beta_h=0$ line have two crossings points with $J=0$ axis, which label the positions of L (at a smaller $h$) and LC (at a larger $h$). With increasing $\tilde{\epsilon}$, the crossing points get closer and merge into one at $\tilde{\epsilon}=\tilde{\epsilon}_1$, signaling the first fixed-point annihilation between L and LC. 

The two crossing points between $\beta_J=0$ and $\beta_h=0$ lines represent KD (at a smaller $h$) and C (at a larger $h$) fixed points respectively. Their annihilation occurs at $\tilde{\epsilon}=\tilde{\epsilon}_2$, 
where
 two curves have a single touching point.  Upon
 further increasing $\tilde{\epsilon}$, these fixed points disappear.

\textit{Role of Berry phase: }
We now discuss
the role of Berry phase and its interplay with Kondo coupling.
From Eq.~\ref{eq:prop}, we see that the Kondo coupling contributes an $i\Omega p(J)$ term to $G_\chi^{-1}$ in the $r=0$ limit. 
It reduces the Berry phase term, $i\Omega$, at finite $J$ and fully removes it at $J=\infty$ and $K=2S$. 
The exact cancellation reveals the nature of Kondo effect which can also be seen by directly solving the problem 
at $J=\infty$ as shown in SM D~\cite{SM}.
It also suggests that, at $J=\infty$, the system should always stay at the Kondo-screened phase. 
Thus, we expect the critical point KD$'$, located at $J=\infty,h=2/(\tilde{\epsilon \pi})$, will be moved to a finite $J$ when
the next-order corrections in $1/S$ 
are included in the RG analysis.

\begin{figure}[b!]
    \centering
    \includegraphics[width=0.5\textwidth]{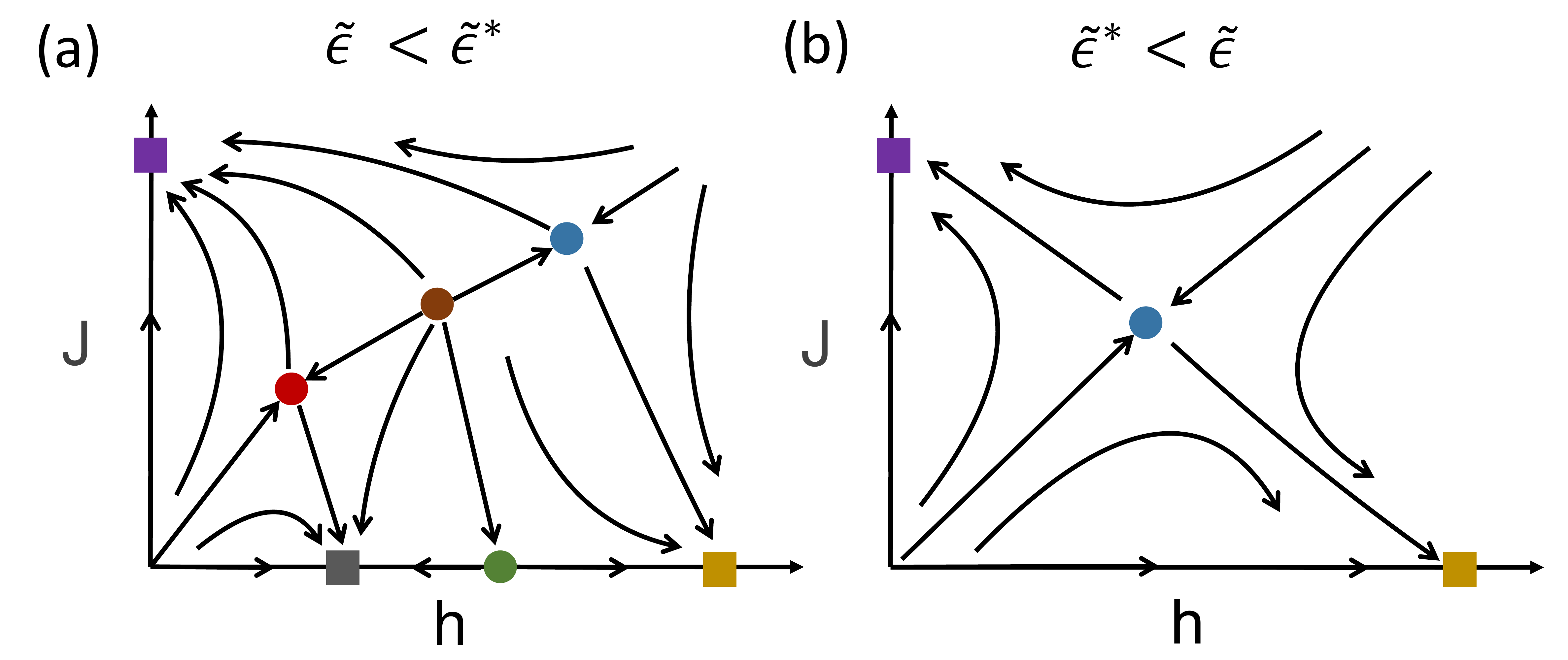}
    \caption{RG flow based on numerical simulations at $S=1/2$ (adapted from Ref.~\cite{Cai_2019}), where $\tilde{\epsilon}^* \simeq 0.265$}
    \label{fig:rg_qmc}
\end{figure}

\textit{Comparison with numerical results:} 
The
BFK model is equivalent to the numerically studied Bose-Fermi Anderson model with a power-law bosonic bath spectral functions
$\rho_b(\omega) =  |\omega|^{1-\epsilon}$
when the Hubbard interaction $U$ is
sufficiently large ~\cite{Cai_2019}. At $S=1/2$, two types of RG flow have been observed in the numerical simulations as shown in Fig.~\ref{fig:rg_qmc} (adapted from Ref.~\cite{Cai_2019}). 
The first one, at $\tilde{\epsilon}<\tilde{\epsilon}^* \simeq 0.265$, is comparable with our analytical results 
at $\tilde{\epsilon} < \tilde{\epsilon}_1$. The other one, at $\tilde{\epsilon}>\tilde{\epsilon}^*$, 
is consistent with our analytical RG flow at $\tilde{\epsilon}>\tilde{\epsilon}_2 $. We note that 
{our analytical analysis has identified}
all the fixed points observed 
numerically
and
the 
values of $\tilde{\epsilon}_{1,2}$ that we have derived turns out to be comparable to
the numerical results.
However, {in further details} there are 
several distinctions in our analytical results.
First, from the analytical calculations
 the KD$'$ fixed point is located at infinite $J$, while it is at 
a finite $J$ numerically.
As we explained before, the next-order corrections (in $1/S$) are expected to move the position of KD$'$ to finite $J$ and we leave
their analysis for a future study. Second, the analytical calculations find a line of fixed points at 
infinite
$h$ instead of a single point as suggested by numerical simulations.
In practice, it's hard to conclude whether L$'$ phase corresponds to a single fixed point or a line of fixed points numerically, 
and it's also possible that the next order 
{(in $1/S$)}
 corrections would shrink the line of fixed points to a single fixed point in our analytical approaches.
We leave the precise
reconcilation
 of this issue for future studies.
 Finally, 
 numerical calculations fail to find 
 RG flows corresponding to $ \tilde{\epsilon}_1<\tilde{\epsilon}<\tilde{\epsilon}_2$. 
 The likely reason for this discrepancy is
  that numerical calculations have so far focused on only a small number of $\tilde{\epsilon}$ 
 and 
 may have missed this narrow region of $\tilde{\epsilon}$. 

\textit{Discussion:} 
Several remarks are in order.
First,
our  results have some general connections with
 the dimensional hierarchy of nonlinear $\sigma$ model 
 with topological terms~\cite{fix_point_annihilation_1,fix_point_annihilation_2,ABANOV2000685}.
An important distinction of our model is the additional coupling to
gapless fermions,
which normalize the effect of the topological term and leads to a richer phase diagram. 

Second,
our model and results are in general relevant to various strongly correlated systems, including heavy fermion metals
and doped Mott-Hubbard systems. In these systems, the 
electrons tend to be localized and form local moments due to the strong Coulomb interactions
{that they experience}. The local moment then interacts with the remaining itinerant electrons and also collective bosonic excitations,
the effects of which are
 captured by
  the BFK model.
 Importantly, the fixed point KD$'$ has a local spin susceptibility $1/|\tau|^{\epsilon}$ \cite{Cai_2019}, which is essential for realizing 
 the EDMFT solutions for the beyond-Landau quantum criticality  \cite{doi:10.7566/JPSJ.83.061005}.
 Thus, the understanding we have reached for the Kondo destruction physics of the SU(2) BFK model 
 is crucially important for understanding
 the quantum criticality of magnetic heavy-fermion metals \cite{QCNP2013,Coleman-Nature,Pas21.1,Kir20.2}.
{Generalizations to including additional local degrees of freedom, such as multipoles \cite{Mar19.1,Liu21.1,Han22.1},
may also be considered.}
Finally, the BFK model also represents an effective description of the doped Mott systems in the form of $t$-$J$-$U$ Hubbard models \cite{SmithSi-edmft}. As such, our results may also play an important role in the understanding of such systems \cite{Fang2022,Bad16.1,Chowdhury2021}.

\textit{Conclusions:}
We have carried out a renormalization group analysis of the spin-isotropic Bose-Fermi Kondo model.
Our RG analysis {determines a set of interacting fixed points with
dynamical Planckian scaling and 
loss of quasiparticles. Our results}
not only provide an understanding of the numerical results on the systematic evolution of the fixed points,
 but also reveals an unexpected 
 {phenomenon
  of sequential
 fixed-point annihilations.}
Our results lay the foundation for the understanding of quantum criticality in
{both}
 spin-isotropic heavy-fermion metals 
 and 
 doped Mott-Hubbard systems.

We thank A. Cai and S. Paschen for useful discussions. This work has primarily been supported by the NSF Grant No. DMR-2220603
 and by the Robert A.\ Welch Foundation Grant No. C-1411.
The work of Q.S.\ was performed in part at the Aspen Center for Physics, which is supported by the NSF grant No.\ PHY-1607611.

\bibliography{bfk_rg}

\clearpage

\onecolumngrid 

\setcounter{figure}{0}
\setcounter{equation}{0}
\makeatletter
\renewcommand{\thefigure}{S\@arabic\c@figure}
\renewcommand{\theequation}{S\arabic{equation}}

\section*{Supplementary Material}

\subsection{A. Renormalization group analysis}
We begin by reiterating that our RG procedure is carried out for the Hamiltonian defined in Eq.~\ref{eq:ham}. 
The effective actions 
$S$ and $S_{\chi}$ are introduced to describe the next-order fluctuations and calculate the correlations functions. 

The relevant correlation functions are:
\ba 
&&\langle \sum_\mu [\phi^\mu(\bm{r},\tau=0)]^2 \rangle _{imp}  \, ,
\nonumber \\
&&\sum_{\sigma,a}\langle c_{\sigma,a}(\bm{r_0},\tau) c_{\sigma,a}^\dag(\bm{r_0},\tau)\rangle _{imp} \, .
\label{eq:correlation}
\ea  
Here, $\langle \cdot \rangle _{imp}$ means we subtract the free-bulk contributions, and the bulk system has spacetime dimension $d=4-\epsilon$. Before performing calculations, we note that the exact formula of above correlations functions are 
\ba 
&&\langle  \sum_\mu \phi^\mu(\bm{r},\tau=0)^2\rangle_{imp}=\int_{\tau,\tau'}\frac{A_\phi}{2\pi } \frac{h}{(\bm{r}^2+\tau^2)^{(2-\epsilon)/2}(\bm{r}^2+\tau'^2)^{(2-\epsilon)/2} }
\langle \sum_\mu S^\mu(\tau) S^\mu(\tau') \rangle_{imp} \nonumber \\
&& -\sum_{\sigma,a}\langle c_{\sigma,a}(\bm{r_0},\tau) c_{\sigma,a}^\dag(\bm{r_0},\tau)\rangle_{imp} = K
\bigg\langle 
\bigg[ 
\delta_{\tau,\tau'}\delta_{k,k'}(\partial_{\tau'}  + \epsilon_{\bk}) 
+ \delta_{\tau,\tau'}J_K\sum_\mu S^\mu(\tau)\sigma^\mu\bigg)
\bigg] 
\bigg\rangle_{imp} -G_0(\bm{r}_0,\tau),
\label{eq:corr_exact}
\ea  
where  
\ba 
h =-4A_\phi g^2\Gamma(-1+\epsilon)\sin(\frac{\epsilon \pi}{2}) \approx  [2\pi A_\phi +O(\epsilon) ]g^2.
\ea
$A_\phi$ is the normalization factor of boson propagator
and $G_0(\bm{r}_0,\tau)$ is the free fermion propagator. Above equations suggest the fermion and boson correlation functions can be represented as functional of $n$-point correlation functions of $S^\mu$ fields. Then, it's sufficient to evaluate the correlation functions of $S^\mu$. To do so, we first derive the effective action of $S^\mu$ by integrating out fermions and bosons:
\ba 
S_{eff}
=-iSS_B 
-K\text{Tr}\log\bigg( \delta_{\tau,\tau'}\delta_{k,k'}(\partial_\tau  + \epsilon_{\bk})- \delta_{\tau,\tau'}J_K\sum_\mu S^\mu(\tau)\frac{\sigma^\mu}{2}\bigg).
\ea 
Using $S_{eff}$ and Eq.~\ref{eq:corr_exact}, we're able to calculate fermion and boson correlation functions. The detailed calculations are shown in Supplementary Material (SM) B and the results are
\ba 
\langle \sum_\mu [ \phi^\mu(\bm{r},\tau=0)]^2\rangle_{imp}
&\propto &  \frac{1}{|\bm{r}|^2}
h\bigg[ 
1- 
\frac{1}{S\pi r}\int_0^J\frac{F(j)}{j}dj +\frac{1}{S\pi \epsilon }[\frac{\log(1+h/c(J)}{c(J)} + \frac{\log(1+h/c(J)^*)}{c(J)^*}]
\bigg] \nonumber \\ 
\sum_{\sigma,a}\langle c_{\sigma,a}(\bm{r_0},\tau) c_{\sigma,a}^\dag(\bm{r_0},\tau)\rangle_{imp} &\propto&
\frac{J^2\text{sgn}(\tau)}{(1+J^2)|\tau|}
\bigg[ 
1 
+ \frac{-1-J^2}{(1+J^2)^2\pi Sr}\int_0^J\frac{F(j)}{j}dj 
+ \frac{1}{(1+J^2)\pi Sr}\int_0^J\frac{2jF(j)}{1+j^2}dj \nonumber\\
&&
-\frac{2}{(1+J^2)r\pi S}
\int_0^J \frac{H(j)}{1+j^2}dj -\frac{-1+J^2}{(1+J^2)^2 \pi S\epsilon}
\bigg( 
\frac{\log(1+h/c(J))}{c(J)}
+
\frac{\log(1+h/c(J)^*)}{c(J)^*}
\bigg) \nonumber \\
&&
+\frac{2iJ}{(1+J^2)^2\pi \epsilon S}
\bigg( -
\frac{\log(1+h/c(J))}{c(J)} 
+
\frac{\log(1+h/c(J)^*)}{c(J)^*}
\bigg) 
\bigg] . \nonumber \\
 \label{eq:corr}
\ea

We are now in the position to carry out renormalization group study. We introduce the following relation between renormalized couplings $J,h$ and bare couplings $J_B,h_B$:
\begin{eqnarray}
&&h_B = Z_h\mu^{\epsilon}h \nonumber \\
&&J_B = Z_J\mu^{r}J 
\label{eq:bare_coupling_supp}
\end{eqnarray}
where $Z_J,Z_h$ are coupling renormalization factor, and $\mu$ is a renormalization energy scale. 
Without loss of generality, we let 
\ba 
Z_h = 1 + \frac{1}{S}x_h \nonumber \\
Z_J = 1 + \frac{1}{S}x_J \nonumber.
\ea 
We determine $x_J$ and $x_h$ by requiring all the poles as a function of $r,\epsilon$  in Eq.~\ref{eq:corr} are minimally removed. This leads to \ba 
x_h &=& \frac{1}{\pi r}\int_0^J\frac{F(j)}{j}dj 
-\frac{1}{\pi \epsilon }
\bigg[ 
\frac{\log(1+h/c(J))}{c(J)}
+
\frac{\log(1+h/c(J)^*)}{c(J)^*}
\bigg] +O(\frac{1}{S})\nonumber\\
x_J&=& \frac{1}{2\pi r}\int_0^J\frac{F(j)}{j}dj 
-\frac{1}{\pi r}\int_0^J\frac{jF(j)}{1+j^2} dj
+\frac{1}{r\pi}\int_0^J\frac{H(j)}{1+j^2}dj +\frac{-1+J^2}{2(1+J^2)\pi \epsilon}\bigg( 
\frac{\log(1+h/c(J))}{c(J)}
+
\frac{\log(1+h/c(J)^*)}{c(J)^*}
\bigg) \nonumber \\
&&
-\frac{iJ}{(1+J^2)\pi \epsilon} 
\bigg( 
-\frac{\log(1+h/c(J))}{c(J)}
+
\frac{\log(1+h/c(J)^*)}{c(J)^*}
\bigg) +O(\frac{1}{S}) \nonumber \\ 
\ea 

Finally, we take $\mu$ derivative of Eq.~\ref{eq:bare_coupling_supp} at fixed $h_B$ and $J_B$ and obtain the following beta functions: 
\ba 
\beta_J &=&  \frac{J-J^3}{2S\pi (1+J^2)}F(J) + \frac{J^2}{S\pi(1+J^2)}H(J) +\bigg[ \frac{-(1-J^2)Jh+2iJ^2h}{2S(1+J^2)\pi c(J)(c(J)+h)}+\text{h.c.}
\bigg] +o(\frac{1}{S}) \nonumber  \, , \\
\beta_h& =& -\frac{1}{S}\tilde{\epsilon} h + \frac{h}{S\pi}F(J)
-\frac{1}{ S\pi }\frac{(c(J)+c(J)^*)h}{|c(J)|^2} 
-\frac{1}{S\pi }\frac{-2h^2-(c(J)+c(J)^*)h}{(c(J)+h)(c(J)^*+h)} +o(\frac{1}{S})  \, .
\ea 
where we also set $r=0$ at the last step.

Note that an alternative way to perform renormalization group study is to start from the action in Eq.~\ref{eq:action} and then expand in powers of $\chi$ fields~\cite{nahum2022fixed}. Combining the quadratic order term $S_{\chi}$ and fourth-order term $S_4$, we could find an interacting field theory of $\chi$: $S_{\chi}+S_4$. Then we are able to perform RG analysis based on this model. The resulting beta functions are expected to be consistent with current approach. 
{We reserve such an analysis for a future study.}

\subsection{B. Calculations of correlation functions}
Here we present the detailed calculation of Eq.~\ref{eq:corr_exact} using the effective action $S_{eff}
$.
We first introduce the spinor representation of $S^\mu$ with $z^\dag \frac{\sigma^\mu}{2} z$ and let $z=z_0+\delta z$ (see Eq.~3). This gives us
\ba 
S^\mu &= & S\delta_{\mu,z} + \delta S^\mu \nonumber \\
\delta S^x &=& \frac{i\sqrt{S}}{\sqrt{2}}(\chi^\dag -\chi) + O(\frac{1}{S^{1/2}})\nonumber \\
\delta S^y &=& \frac{\sqrt{S}}{\sqrt{2}}(\chi^\dag +\chi) + O(\frac{1}{S^{1/2}}) \nonumber  \\
\delta S^z &=& -|\chi|^2+O(\frac{1}{S}) 
. 
\label{eq:s_exp}
\ea 
We expand the action in Eq.~\ref{eq:action} to the quadratic order of $\chi$ fields:
\ba 
S_{eff}
&=& S_0 +S_\chi \nonumber \\
S_0&=&-\frac{Sg^2}{4}\int_{\tau,\tau'} \frac{\sum_\mu(z_0^\dag \sigma^\mu z_0)^2}{2}G_\phi(\tau-\tau')
-K\text{Tr}\log\bigg( 
\delta_{\tau,\tau'}\delta_{k,k'}(\partial_{\tau'}  + \epsilon_{\bk}) 
+ \delta_{\tau,\tau'}J_K\sum_\mu (z_0^\dag \sigma^\mu z_0)\sigma^\mu\bigg) ,
\label{eq:simp_exp}
\ea 
with $S_0$ the saddle point contribution at order of $S$ and $S_\chi$ describing fluctuations at order $S^0$. We now derive the formula of $S_\chi$. It has contribution from three part: the Berry phase, bosonic coupling and Kondo
interaction.
The Berry phase term 
gives  
\ba 
\int i\Omega|\chi(\Omega)|^2 \frac{d\Omega}{2\pi} \, ,
\label{eq:berry_chi}
\ea 
{while}
 the bosonic coupling 
yields
\ba 
&&-\frac{g^2}{2}\int_{\tau,\tau'}G_\phi(\tau-\tau')[\chi^\dag(\tau)\chi(\tau') + \chi(\tau)\chi^\dag(\tau') -|\chi(\tau)|^2 -|\chi(\tau')|^2] \nonumber \\
&=&\int h|\Omega|^{1-\epsilon} |\chi(i\Omega)|^2 \frac{d\Omega}{2\pi} \, ,
\label{eq:bose_chi}
\ea 
and, finally, the
 Kondo coupling gives (by expanding the fermion determinant)
\ba 
&&\frac{\kappa J_K^2}{4} \int_{\tau,\tau'}\bigg[ G_s(\tau-\tau')G_s(\tau'-\tau) -G_a(\tau-\tau')G_a(\tau'-\tau)\bigg]
\bigg[
\chi^\dag(\tau)\chi(\tau') + \chi^\dag(\tau')\chi(\tau) - |\chi(\tau)|^2 - |\chi(\tau')|^2
\bigg]\nonumber \\
&&+\frac{\kappa J_K^2}{2}\int_{\tau,\tau'}\bigg[ 
G_s(\tau-\tau')G_a(\tau'-\tau)
\bigg] 
\bigg[ 
-\chi^\dag(\tau)\chi(\tau') + \chi(\tau)\chi^\dag(\tau')
\bigg] \nonumber \\
&=&\int \frac{2\kappa J^2 |\Omega|^{1-2r} }{\pi(1+J^2 |\Omega|^{-2r})^2}|\chi(i\Omega)|^2 \frac{d\Omega}{2\pi}+\int \frac{i\Omega \kappa }{\pi }
\bigg[ 
\frac{J|\Omega|^{-r} -J^3 |\Omega|^{-3r}}{(1+J^2|\Omega|^{-2r})^2} - \arctan(J|\Omega|^{-r})
\bigg] |\chi(i\Omega)|^2 \frac{d\Omega}{2\pi} ,
\label{eq:fermi_chi}
\ea 
where $G_s(\tau) = \sum_\sigma G_\sigma(\tau)$ and $G_a(\tau) = \sum_\sigma \sigma G_\sigma(\tau)$ and $G_\sigma(\tau)$ denotes the conduction electron Green's function at leading order $(S^1)$. The explicit formula of these functions are
\ba 
G_s(i\omega) &= \frac{-2i\rho_0\pi |\omega|^{-r}}{1+\rho_0^2\pi^2J_K^2 |\omega|^{-2r}} \text{sgn}(\omega) \nonumber \\
G_a(i\omega) &= \frac{-2\rho_0^2\pi^2J_K |\omega|^{-2r}}{1+\rho_0^2\pi^2J_K^2 |\omega|^{-2r}} .
\ea 
In addition, the coefficient before each $|\Omega|^{-r}$ should be a function of $r$. However, we only keep the leading order term ($r^0$) since we would set $r=0$ in the final step. Combining Eqs.~\ref{eq:berry_chi}, \ref{eq:bose_chi} and ~\ref{eq:fermi_chi}, we have the quadratic theory of the $\chi$ fields:
\ba 
&&S_{\chi} = \int  G^{-1}_\chi(i\Omega)|\chi(i\Omega)|^2 \frac{d\Omega}{2\pi} \nonumber \\
&&G_\chi^{-1}(i\Omega) = i\Omega \bigg[ 1+p(J|\Omega|^{-r})\bigg] +|\Omega|\bigg[ q(J|\Omega|^{-r}) + h|\Omega|^{-\epsilon}\bigg].
\label{eq:Schisupp}
\ea 

For future reference, we introduce the series expansion of $G_\chi(i\Omega)$
\ba 
G_\chi(i\Omega) &=& \frac{1}{|\Omega|} 
\frac{1}{i(\text{sgn}(\Omega)+p(J|\Omega|^{-r})) + q(J|\Omega|^{-r}) 
} + \frac{1}{|\Omega|}\sum_{m=1}^\infty \frac{(-h)^m|\Omega|^{-m \epsilon} }
{[i(\text{sgn}(\Omega) +p(J)) + q(J)]^{m+1}}  
\ea 
Since we set $r=0$ in the last step, we drop the $r$ dependence in the summation by setting $r=0$. Including this contribution won't change the beta functions at the leading order. We can further expand the real and imaginary part of $G_\chi(i\Omega)$ as the following 
\ba 
&&G_\chi(i\Omega) + G_\chi(-i\Omega) = 
\sum_{n=0}^\infty a_n J^n |\Omega|^{-1-nr} + \sum_{m=1}^\infty 
(-h)^m \bigg(\frac{1}{(c(J)^{*})^{m+1}}+\frac{1}{c(J)^{m+1}}\bigg) 
|\Omega|^{-1-m\epsilon} \nonumber \\
&&G_\chi(i\Omega)-G_\chi(i\Omega) = 
\sum_{n=0}^\infty ib_n J^n|\Omega|^{1-nr} +   \sum_{m=1}^\infty 
(-h)^m \bigg(-\frac{1}{(c(J)^{*})^{m+1}}+\frac{1}{c(J)^{m+1}}\bigg) 
|\Omega|^{1-m\epsilon}
\ea 
where $a_n, b_n$ are the Taylor coefficients of $F(j)$ and $H(j)$ at $j=0$ and 
\ba 
F(j) &=& \frac{2q(j)}{(1+p(j))^2+q(j)^2}\nonumber \\ 
H(j) &= &\frac{-2(1+p(j))}{(1+p(j))^2 + q(j)^2}\nonumber \\
c(J) &= & i(1+p(J)) + q(J) \nonumber .
\ea 

Now, we combine Eqs.~\ref{eq:corr_exact}, \ref{eq:s_exp}, \ref{eq:simp_exp} and \ref{eq:Schisupp} and calculate the correlation functions of fermions and bosons. 
We first expand Eq.~\ref{eq:corr_exact} using Eq.~\ref{eq:s_exp}. For bosons, we have
\ba 
&&\langle  \sum_\mu \phi^\mu(\bm{r},\tau=0)^2\rangle_{imp}\nonumber \\ &=&\int_{\tau,\tau'}\frac{A_\phi}{2\pi } \frac{h}{(\bm{r}^2+\tau^2)^{(2-\epsilon)/2}(\bm{r}^2+\tau'^2)^{(2-\epsilon)/2} }S^2 \nonumber \\
&&+ 
\int_{\tau,\tau'}\frac{A_\phi}{2\pi } \frac{h}{(\bm{r}^2+\tau^2)^{(2-\epsilon)/2}(\bm{r}^2+\tau'^2)^{(2-\epsilon)/2} }\langle -S\delta S^z(\tau) -S\delta S^z(\tau') + \delta S^x(\tau) \delta S^x(\tau') + \delta S^y(\tau) \delta S^y(\tau') \rangle_{S_\chi}  + O(S^0) \nonumber \\
\ea 
where the last line denotes the contribution of the next-order fluctuations and are evaluated with respect to $S_{\chi}$. 
For the fermionic fields, it gives 
\ba 
&&-\langle  \sum_{a,\sigma} c_{\bm{r_0},a,\sigma}(\tau)
c_{\bm{r_0},a,\sigma}^\dag(0)\rangle \nonumber \\
&=&K\bigg[G_s(\tau) -G_0(\tau) \nonumber \\
&&
+\int_{\tau_1}\sum_\mu \frac{J_K\text{Tr}[G(\tau-\tau_1) \sigma^\mu G(\tau_1)]}{S}
 \langle \delta S^\mu(\tau_1)\rangle_{S_{\chi}}
+\int_{\tau_1,\tau_2}\sum_{\mu,\nu}\frac{J^2_K\text{Tr}[G(\tau-\tau_1)  \sigma^\mu G(\tau_1-\tau_2) \sigma^\nu G(\tau_2)]}{S^2} 
\langle
\delta S^\mu(\tau_1) \delta S^\nu(\tau_2)
\rangle_{S_{\chi}}
\bigg]\nonumber \\
&&+ O(S^0)
\ea 
where the third line denotes the contribution of the next-order fluctuations and are evaluated with respect to $S_{\chi}$ and $G(\tau) = \text{diag}[\frac{G_s(\tau)+G_a(\tau)}{2},\frac{G_s(\tau)-G_a(\tau)}{2}]$.

To get the final results, we only need to calculate the correlation functions of $\delta S^\mu$ (or $\chi$) that appearing in the above equations. The boson correlation function becomes
\ba 
&&\langle  \sum_\mu \phi^\mu(\bm{r},\tau=0)^2\rangle_{imp}\nonumber \\
&=&\int_{\tau,\tau'}\frac{A_\phi}{2\pi } \frac{h}{(\bm{r}^2+\tau^2)^{(2-\epsilon)/2}(\bm{r}^2+\tau'^2)^{(2-\epsilon)/2} }
S^2 \nonumber\\
&&\bigg[ 
1 + \frac{1}{S}\langle \chi^\dag(\tau)\chi(\tau') + \chi^\dag (\tau')\chi(\tau') - |\chi(\tau)|^2 -|\chi(\tau')|^2\rangle 
\bigg]+O(S^0)\nonumber \\
&=&\frac{A_\phi \pi S^2}{2|\bm{r}|^2}h 
+  
\frac{A_\phi Sh}{2\pi }
\int_0^\infty 
2(G_\chi(\Omega)+G_\chi(-\Omega))
\bigg( 
2\pi x^{-(1-\epsilon)}B[\frac{\epsilon-1}{2},x]^2|\Omega|^{2-2\epsilon} - \frac{\pi^2}{|\bm{r}|^2}
\bigg) 
\frac{d\Omega}{2\pi} \nonumber +O(S^0)\\
&=&\frac{A_\phi \pi S^2}{2|\bm{r}|^2}h 
+  
\frac{A_\phi Sh}{2\pi } 
\int_0^\infty 
2\bigg[ 
\sum_{n=0}^\infty  a_nJ^n|\Omega|^{1-nr}
+\sum_{m=1}^\infty (-h)^m \bigg(\frac{1}{(c(J)^{*})^{m+1}}+\frac{1}{c(J)^{m+1}}\bigg) 
|\Omega|^{-1-m\epsilon} 
\bigg] \nonumber \\
&&\frac{1}{|\bm{r}|^{2}}
\bigg(
2\pi x^{1-\epsilon}B[\frac{\epsilon-1}{2},x]^2-
\pi 2^{-\epsilon}\Gamma((1-\epsilon)/2)^2
\bigg) 
\frac{d\Omega}{2\pi} +O(S^0)\nonumber \\
&=&
\frac{A_\phi \pi S^2}{2|\bm{r}|^2}h 
+  
\frac{A_\phi S h}{\pi |\bm{r}|^2} \bigg[ \frac{1}{r}
\sum_{n=0}^\infty \frac{-a_nJ^n\pi}{2n} 
+\frac{1}{\epsilon}\sum_{m=1}(-h)^m 
\bigg(\frac{1}{(c(J)^{*})^{m+1}}+\frac{1}{c(J)^{m+1}}\bigg) \frac{-\pi}{2n\epsilon} 
+O(r^0,\epsilon^0)\bigg] +O(S^0) \nonumber \\
&=&\frac{A_\phi \pi S^2}{2|\bm{r}|^2}
h\bigg\{ 
1- \frac{1}{S}\bigg[\frac{1}{\pi r}\int_0^J\frac{F(j)}{j}dj 
+\frac{1}{\pi \epsilon }[\frac{\log(1+h/c(J)}{c(J)} + \frac{\log(1+h/c(J)^*)}{c(J)^*}]
+O(\epsilon^0,r^0)\bigg]+o(\frac{1}{S})
\bigg\} 
\label{eq:chi_prop}
\ea 
where $x=|\Omega||\bm{r}|$ and $B(a,x)$ is the modified Bessel functions of the second kind. In the above expression, we only keep track of the poles as a function of $\epsilon$ and $r$. We also drop the $\frac{r}{\epsilon}$ poles which goes to zero as we set $r=0$. 

For the fermion correlation function,
we first perform Fourier transformation and separate the next-order contributions into four parts $I_1,I_2,I_3,I_4$:
\ba 
G_f(i\omega) = -\int_\tau\langle  \sum_{a,\sigma} c_{\bm{r_0},a,\sigma}(\tau)
c_{\bm{r_0},a,\sigma}^\dag(0)\rangle_{imp}
e^{i\omega\tau} = K\bigg[ \frac{2i\pi \rho_0 J^2}{1+J^2} \text{sgn}(\omega)
+I_1+I_2+I_3+I_4+o(\frac{1}{S})\bigg] 
\ea 
Each part is shown below:
\begin{eqnarray*}
I_1 &=& 
\frac{J_K^2}{4S}(G_s(i\omega)^2 +G_a(i\omega)^2)
\int G_\chi(i\Omega) (G_s(i\omega+i\Omega)+G_s(i\omega-i\Omega)-G_s(i\omega)-G_s(i\omega))\frac{d\Omega}{2\pi} \\
&=&\frac{-2i\rho_0\pi J^2(-1+J^2)}{(1+J^2)^2S}
\int_0^\infty \bigg[ \sum_{n=0}^\infty a_nJ^n|\Omega|^{-1-nr} +\sum_{m=1}^{\infty}(-h)^m \bigg(\frac{1}{(c(J)^{*})^{m+1}}+\frac{1}{c(J)^{m+1}}\bigg) 
|\Omega|^{-1-m\epsilon}\bigg]\\
&&\sum_{p=0}^\infty (-J^2)^p
\bigg[ \text{sgn}(\omega+\Omega)|\omega+\Omega|^{-2pr-r} +
 \text{sgn}(\omega-\Omega)|\omega-\Omega|^{-2pr-r}
 -2
  \text{sgn}(\omega)|\omega|^{-2pr-r}
\bigg]\frac{d\Omega}{2\pi} +O(r^0)\\
&=&\frac{-i\rho_0\pi J^2(-1+J^2)}{(1+J^2)^2\pi S}\bigg[ \sum_{n=0}^\infty a_nJ^n\sum_{p=0}^\infty (-J^2)^P\frac{-2}{nr} 
+
\sum_{m=1}^\infty (-h)^m \bigg(\frac{1}{(c(J)^{*})^{m+1}}+\frac{1}{c(J)^{m+1}}\bigg) \frac{-2}{m\epsilon} 
\bigg] \text{sgn}(\omega)  \\
&=&\frac{2i\rho_0\pi J^2(-1+J^2)}{(1+J^2)^3\pi S}
\bigg[ 
\frac{1}{r} \int_0^J \frac{F(j)}{j}dj
-\frac{1}{\epsilon}\bigg( 
\frac{\log(1+h/c(J))}{c(J)}
+
\frac{\log(1+h/c(J)^*)}{c(J)^*}
+O(r^0,\epsilon^0)\bigg) 
\bigg] \text{sgn}(\omega) .
\end{eqnarray*}

\begin{eqnarray*}
I_2 &=& 
-\frac{J_K^2}{2S}G_s(i\omega)G_a(i\omega)
\int G_\chi(i\Omega)
(G_a(i\omega+i\Omega)+G_a(i\omega-i\Omega)-G_a(i\omega)-G_a(i\omega))\frac{d\Omega}{2\pi} \\
&=&-\frac{2iJ^3(-2J\rho_0\pi)}{(1+J^2)^2S}\text{sgn}(\omega)
\int_0^\infty 
\bigg[ 
\sum_{n=0}^\infty a_nJ^n|\Omega|^{-1-nr} +\sum_{m=1}^{\infty}(-h)^m \bigg(\frac{1}{(c(J)^{*})^{m+1}}+\frac{1}{c(J)^{m+1}}\bigg) 
|\Omega|^{-1-m\epsilon}
\bigg] \\
&&
\sum_{p=0}^\infty
(-J^2)^p \bigg[ 
|\omega+\Omega|^{-2pr-2r}
+|\omega-\Omega|^{-2pr-2r}
-2|\omega|^{-2pr-2r}
\bigg] 
\frac{d\Omega}{2\pi}\\
&=&-\frac{2iJ^3(-2J\rho_0\pi)}{2\pi (1+J^2)^2S}\text{sgn}(\omega)
\bigg[ 
\sum_{n=0}^\infty a_n
\sum_{p=0}^\infty (-J^2)^p
J^n(\frac{2}{2pr+2r+nr} - \frac{2}{nr})
+O(r^0,\epsilon^0)
\bigg] 
 \\
&=& \frac{2iJ^4\rho_0\pi}{\pi (1+J^2)^2S}\text{sgn}(\omega)
 \bigg[ 
 \frac{1}{rJ^2}\int_0^J \frac{2jF(j)}{1+j^2}dj 
 -\frac{2}{r(1+J^2)}\int_0^J\frac{F(j)}{j}dj
 +O(r^0,\epsilon^0)\bigg].
\end{eqnarray*}

\begin{eqnarray*}
I_3 &=&\frac{J_K^2}{4S}
\bigg[ 
G_a(i\omega)^2 +G_s(i\omega)^2 
\bigg] \int G_\chi(i\Omega) [
G_a(i\Omega+i\omega) - G_a(i\omega-i\Omega)]
\frac{d\Omega}{2\pi} \\
&=& \frac{J^2(-1+J^2)}{2\pi(1+J^2)^2S} 
\int_0^\infty \bigg[ 
\sum_{n=0}ib_nJ^n|\Omega|^{1-nr} +\sum_{m=1}(-h)^m
\bigg(-\frac{1}{(c(J)^{*})^{m+1}}+\frac{1}{c(J)^{m+1}}\bigg)|\Omega|^{-1-m\epsilon} \bigg]
\\
&&\bigg[ -2J\rho_0\pi \sum_{p=0}^\infty (-J^2)^p 
\bigg( 
|\Omega+\omega|^{-2r-2pr} -
|-\Omega+\omega|^{-2r-2pr}
\bigg) 
\bigg] 
d\Omega \\
&=&\frac{1}{S}[0+O(r^0,\epsilon^0)].
\end{eqnarray*} 

\begin{eqnarray*} 
I_4 &=& 
\frac{-J_K^2}{2}G_a(i\omega)G_s(i\omega)
\int G_\chi(i\Omega) \bigg(
G_s(i\Omega+i\omega)-G_s(i\omega-i\Omega)
\bigg) \frac{d\Omega}{2\pi} \\
&=&\frac{-2J^3\rho_0\pi}{S(1+J^2)^2}\int_0^\infty 
\bigg[ 
\sum_{n=0}ib_nJ^n|\Omega|^{1-nr} +\sum_{m=1}(-h)^m
\bigg(-\frac{1}{(c(J)^{*})^{m+1}}+\frac{1}{c(J)^{m+1}}\bigg)|\Omega|^{-1-m\epsilon} \bigg]\\
&&
\sum_{p=0}(-J^2)^p
\bigg[ 
\text{sgn}(\omega+\Omega)|\omega+\Omega|^{-2pr-r}
-\text{sgn}(\omega-\Omega)|\omega-\Omega|^{-2pr-r}
\bigg] 
\frac{d\Omega}{\pi} \\
&=&\frac{-2J^3\rho_0\pi}{S(1+J^2)^2} 
\bigg[ 
\sum_{n=0}ib_nJ^n\sum_{p=0} (-J^2)^p \frac{2}{\pi(nr+2pr+r)} +\sum_{m=1}(-h)^m\frac{1}{1+J^2}
\bigg(-\frac{1}{(c(J)^{*})^{m+1}}+\frac{1}{c(J)^{m+1}}\bigg)\frac{2}{\pi m\epsilon} \bigg]\text{sgn}(\omega)  \\
&=&\frac{-2J^3\rho_0\pi}{(1+J^2)^2} 
\bigg[ \frac{2}{ r\pi J}\int_0^J\frac{iH(j)}{(1+j^2)}dj
+\frac{2}{\pi \epsilon(1+J^2)}\bigg(
-\frac{\log(1+h/c(J))}{c(J)}
+\frac{\log(1+h/c(J)^*)}{c(J)^*}
\bigg) +O(r^0,\epsilon^0)
\bigg]\text{sgn}(\omega)
.
\\
\end{eqnarray*}
In the above derivation, we only keep terms that diverge as $r\rightarrow 0$ and $\epsilon \rightarrow 0$, with $r$ going to zero first.

Summing over all the contributions, we have 
\ba 
&&G_f(i\omega) \nonumber \\
&=& 2iK\pi \rho_0\text{sgn}(\omega) \frac{J^2}{1+J^2} 
\bigg[ 
1
+ \frac{-1+J^2}{(1+J^2)^2\pi Sr}\int_0^J\frac{F(j)}{j}dj 
+ \frac{1}{(1+J^2)\pi Sr}\int_0^J\frac{2jF(j)}{1+j^2}dj \nonumber\\
&& -\frac{2J^2}{(1+J^2)^2\pi Sr}
\int_0^J\frac{F(j)}{j}dj
-\frac{2}{(1+J^2)r\pi S}
\int_0^J \frac{H(j)}{1+j^2}dj \nonumber \\ 
&&-\frac{-1+J^2}{(1+J^2)^2 \pi S\epsilon}
\bigg( 
\frac{\log(1+h/c(J))}{c(J)}
+
\frac{\log(1+h/c(J)^*)}{c(J)^*}
\bigg) 
+\frac{2iJ}{(1+J^2)^2\pi \epsilon S}
\bigg( -
\frac{\log(1+h/c(J))}{c(J)}
+
\frac{\log(1+h/c(J)^*)}{c(J)^*}
\bigg) 
\bigg] 
\nonumber \\
\label{eq:g_prop}
\ea 

In summary, we have 
\ba 
\langle \sum_\mu [ \phi^\mu(\bm{r},\tau=0)]^2\rangle_{imp}
&\propto &  \frac{1}{|\bm{r}|^2}
h\bigg[ 
1- 
\frac{1}{S\pi r}\int_0^J\frac{F(j)}{j}dj +\frac{1}{S\pi \epsilon }[\frac{\log(1+h/c(J)}{c(J)} + \frac{\log(1+h/c(J)^*)}{c(J)^*}]
\bigg] \nonumber \\ 
\sum_{\sigma,a}\langle c_{\sigma,a}(\bm{r_0},\tau) c_{\sigma,a}^\dag(\bm{r_0},\tau)\rangle_{imp} &\propto&
\frac{J^2\text{sgn}(\tau)}{(1+J^2)|\tau|}
\bigg[ 
1 
+ \frac{-1-J^2}{(1+J^2)^2\pi Sr}\int_0^J\frac{F(j)}{j}dj 
+ \frac{1}{(1+J^2)\pi Sr}\int_0^J\frac{2jF(j)}{1+j^2}dj \nonumber\\
&&
-\frac{2}{(1+J^2)r\pi S}
\int_0^J \frac{H(j)}{1+j^2}dj -\frac{-1+J^2}{(1+J^2)^2 \pi S\epsilon}
\bigg( 
\frac{\log(1+h/c(J))}{c(J)}
+
\frac{\log(1+h/c(J)^*)}{c(J)^*}
\bigg) \nonumber \\
&&
+\frac{2iJ}{(1+J^2)^2\pi \epsilon S}
\bigg( -
\frac{\log(1+h/c(J))}{c(J)} 
+
\frac{\log(1+h/c(J)^*)}{c(J)^*}
\bigg) 
\bigg] . \nonumber 
\ea

\subsection{C. Beta function analysis} 
\begin{figure*}
    \centering
    \includegraphics[width=\textwidth]{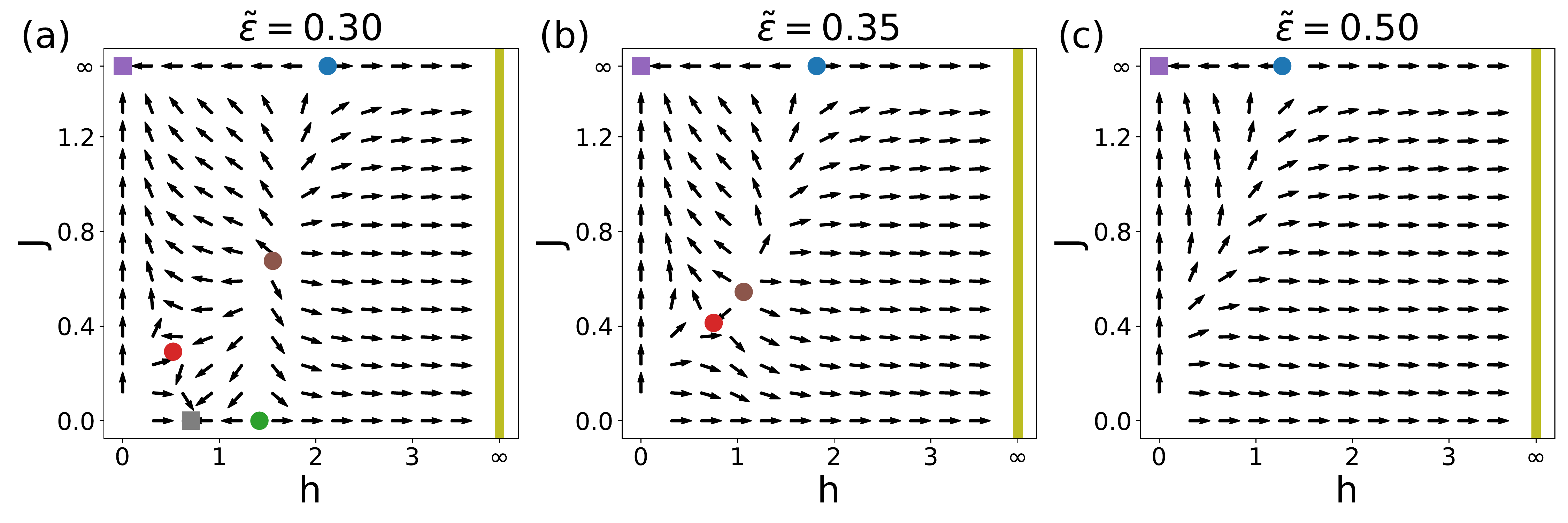}
    \caption{Renormalization group flow at $\tilde{\epsilon}<\tilde{\epsilon}_1$ (a), $\tilde{\epsilon}_1<\tilde{\epsilon}<\tilde{\epsilon}_2$ (b) and $\tilde{\epsilon}_2<\tilde{\epsilon}$ (c) obtained from numerical evaluation. For purpose of illustration, we normalize all the vectors to have the same length. }
    \label{fig:rg_num} 
\end{figure*}

We numerically evaluate the beta functions and the resulting RG flows are shown in Fig.~\ref{fig:rg_num}.
Besides numerical evaluations, we analyze the beta equations in certain extreme cases.

At $h=0$, the model becomes a standard Kondo impurity model. The beta function of $J$ is 
\ba 
\beta_J\bigg|_{h=0} &=& \frac{J-J^3}{2S\pi (1+J^2)}F(J) + \frac{J^2}{S\pi(1+J^2)}H(J)  \nonumber \\
&=&-\frac{2J^2}{S\pi} +\frac{2\kappa J^3}{S\pi^2}  +O(J^4/S). 
\ea
Numerically, we find $J$ is always relevant at $\kappa=2$ and the system flows to a strong coupling fixed point $K$ at $J=\infty$. The first two terms in the second line are also consistent with the result obtained from perturbation theory as shown in Ref.~\cite{affleck1995conformal} (note that we've scaled the Kondo coupling).

At $J=0$, the model becomes a Bose-Kondo model with the following beta function of $h$:
\ba 
\beta_h\bigg|_{J=0} = -\frac{1}{S}\tilde{\epsilon} h + \frac{2h^2}{S\pi (1+h^2)}.
\ea 
This is consistent with previous results obtained in Refs.~\cite{cuomo2022spin,nahum2022fixed}. It gives two stable fixed points at $h = (1-\sqrt{1-\tilde{\epsilon}^2})/ {\tilde{\epsilon}},\infty$ (L and L$'$) and one critical point at $h=(1+\sqrt{1-\tilde{\epsilon}^2})/ {\tilde{\epsilon}}$ (LC). After introducing the Kondo coupling, we find $J$ is irrelevant at both L and LC, but marginal at L$'$. 

At $h=\infty$, we have a line of fixed points (L$'$) where $J$ is always marginal. The beta function at the large $h$ is:
\ba 
\beta_J = \frac{J-J^3}{\pi S (1+J^2)}\frac{1}{h} + O(\frac{1}{h^2S}).
\ea 
Clearly, $\beta_J \rightarrow 0 $ as $h\rightarrow \infty$.

At $J=\infty$, besides the K fixed point, there is an unstable fixed point KD$'$ at $(h_c=\frac{2}{\tilde{\epsilon}\pi},J=\infty)$. To see the existence of such a critical point, we give the beta function of $h$ at large $J$:
\ba 
\beta_h= -\frac{\tilde{\epsilon} h}{S}+\frac{2}{S\pi} + O(\frac{1}{SJ}). 
\ea 
It suggests that $h$ flows to $0$ ($\infty$) when $h<(>) h_c$ at $J\rightarrow \infty$. 

\subsection{D. Topological contribution from the Kondo coupling} 
In this section, we'll show that the Kondo coupling 
contributes an additional topological term to the impurity spin at sufficiently large values of $J_K$. For the purpose of discussion, we consider pure Kondo model, in other words, the BFK model at $g=0$ or $h=0$. We then integrate out the fermionic fields, which gives the following action
\ba 
S_{K}
=-iSS_B 
-K\text{Tr}\log\bigg( \delta_{\tau,\tau'}\delta_{k,k'}(\partial_\tau  + \epsilon_{\bk})- \delta_{\tau,\tau'}J_K\sum_\mu S^\mu(\tau)\frac{\sigma^\mu}{2}\bigg).
\ea 
We now evaluate the fermion determinant contribution at sufficiently large $J_K$ via  a Trotter decomposition
\ba 
\exp\bigg ( K\text{Tr}\log\bigg( \delta_{\tau,\tau'}\delta_{k,k'}(\partial_\tau  + \epsilon_{\bk})- \delta_{\tau,\tau'}J_K\sum_\mu S^\mu(\tau)\frac{\sigma^\mu}{2}\bigg) \bigg) = \bigg[\lim_{N\rightarrow \infty }\text{Tr}[\prod^N_i \exp[-H(\tau_i) d\tau]]\bigg]^K
\ea 
where
\ba 
H(\tau_i) = J_K\sum_\mu c_{\bm{r}_0}\frac{\sigma^\mu}{2} c_{\bm{r}_0}S^\mu(\tau_i) + \sum_{\bk \sigma } \epsilon_{\bk}c_{\bk,\sigma}^\dag c_{\bk,\sigma}.
\ea 
Since the $K$ orbitals are
degenerate, it's sufficient to consider one orbital and multiply the contribution by $K$. When $J_K$ is sufficiently large, the hopping between site $\bm{r}_0$ and the other sites is a high order effect ($\frac{t^2}{J_K}$). Here, we focus on the leading order contribution, and it's sufficient to take $H(\tau_i) = J_K/2\sum_\mu c_{\bm{r}_0}\sigma^\mu c_{\bm{r}_0}S^\mu(\tau_i)$. The eigenstates and eigenvalues of $H(\tau_i)$ are 
\ba 
&&|v_{+}(\tau_i)\rangle = \cos(\frac{\theta(\tau_i)}{2})c_{\bm{r_0},\up}^\dag 
|0\rangle +\sin(\frac{\theta(\tau_i)}{2})e^{i\phi(\tau_i)}c_{\bm{r_0},\dn}^\dag |0\rangle \nonumber \\
&&|v_{-}(\tau_i)\rangle = -\sin(\frac{\theta(\tau_i)}{2})c_{\bm{r_0},\up}^\dag 
|0\rangle +\cos(\frac{\theta(\tau_i)}{2})e^{i\phi(\tau_i)}c_{\bm{r_0},\dn}^\dag |0\rangle \nonumber \\
&&|v_0(\tau_i)\rangle = |0\rangle \nonumber  \\
&&|v_d(\tau_i)\rangle = |\up\dn\rangle \nonumber \\
&&E_{\pm}(\tau_i) = \pm \frac{J_KS}{2},\quad E_0(\tau_i) = 0 , \quad E_d(\tau_i) =0
\ea 
where we let $\bm{S}(\tau_i) =S (\sin(\theta(\tau_i))\cos(\phi(\tau_i)), \sin(\theta(\tau_i))\sin(\phi(\tau_i)), \cos(\theta(\tau_i)))$. We have the following decomposition via the eigenvalues and eigenstates
\ba 
e^{-H(\tau_i)d\tau} &=& e^{-E_+(\tau_i)d\tau}|v_{+}(\tau_i)\rangle\langle v_{+}(\tau_i)|+ e^{-E_-(\tau_i)d\tau}|v_{-}(\tau_i)\rangle\langle v_{-}(\tau_i)|
\nonumber \\
&&
+e^{-E_0(\tau_i)d\tau} |v_0(\tau_i)\rangle \langle v_0(\tau_i)| 
+e^{-E_d(\tau_i)d\tau} |v_d(\tau_i)\rangle \langle v_d(\tau_i)|
.
\ea 
Combining the above identity with a Trotter decomposition of the fermion determinant, we obtain the following leading order contribution
\ba 
&&\exp(K\text{Tr}\log\bigg( \delta_{\tau,\tau'}\delta_{k,k'}(\partial_\tau  + \epsilon_{\bk})- \delta_{\tau,\tau'}J_K\sum_\mu S^\mu(\tau)\frac{\sigma^\mu}{2}\bigg)) \nonumber \\
&\propto& \exp\bigg(K\int_\tau \langle \partial_\tau v_-(\tau)|v_-(\tau)\rangle d\tau +   O(\frac{1}{J_K}) \bigg)\nonumber \\
&\propto& \exp\bigg(i\frac{K}{2}\int_\tau (1-\cos(\theta))\partial_\tau \phi d\tau +   O(\frac{1}{J_K}) \bigg)\nonumber \\
&=& \exp\bigg( -i\frac{K}{2}S_B+
O(\frac{1}{J_K})\bigg).
\ea 
The effective action  then
becomes 
\ba 
S_{K}
= -i(S-\frac{K}{2})S_B + O(\frac{1}{J_K})
\ea 
which indicates the non-trivial topological contribution from the coupling to the gapless fermions. 
Qualitatively,
 the spin has been Kondo-screened, with the effective spin becoming $S-\frac{K}{2}$. 
 {For the}
 perfect screening case
 {that we consider}, $S-\frac{K}{2}=0$, and we obtain a Kondo singlet.

\end{document}